# The BepiColombo Solar Conjunction Experiments Revisited

Ivan di Stefano[1], Paolo Cappuccio[1], and Luciano Iess[2]
*Sapienza University of Rome, Rome, 00184, Italy*

**BepiColombo ESA/JAXA mission is currently in its 7-year cruise phase towards Mercury. The Mercury Orbiter Radioscience Experiment (MORE), one of the 16 experiments of the mission, will start its scientific investigation during the superior solar conjunction (SSC) in March 2021 with a test of general relativity (GR). Other solar conjunctions will follow during the cruise phase, providing several opportunities to improve the results of the first experiment. MORE radio tracking system allows to establish precise ranging and Doppler measurements almost at all solar elongation angles (up to 7-8 solar radii), thus providing an accurate measurement of the relativistic time delay and frequency shift experienced by a radio signal during a SSC. The final objective of the experiment is to place new limits to the accuracy of the GR as a theory of gravity in the weak-field limit. As in all gravity experiments, non-gravitational accelerations acting on the spacecraft are a major concern. Because of the proximity to the Sun, the spacecraft will undergo severe solar radiation pressure acceleration, and the effect of the random fluctuations of the solar irradiance may become a significant source of spacecraft buffeting. In this paper we address the problem of a realistic estimate of the outcome of the SSC experiments of BepiColombo, by including in the dynamical model the effects of random variations in the solar irradiance. We propose a numerical method to mitigate the impact of the variable solar radiation pressure on the outcome of the experiment. Our simulations show that, with different assumptions on the solar activity and observation coverage, the accuracy obtainable on the estimation of $\gamma$ lays in the range $[6\text{-}13] \times 10^{-6}$.**

---

[1] PhD student, Department of Mechanical and Aerospace Engineering.
[2] Full professor, Department of Mechanical and Aerospace Engineering

# I. Introduction

The Newtonian view of an absolute space [1] as a structure with its own properties was sharply criticized already in the 18th century by George Berkeley [2]. The picture outlined by Mach [3] states that the motion of a particle would be meaningless if not considered relative to other matter in the universe. Although general relativity (GR) creates a connection between the geometrical properties of the space-time and the mass-energy in it, some solution of the GR field equations exists even without the presence of mass [4, 5, 6].

One of the possible interpretations of this, is that there are some physical cases which cannot be correctly described by Einstein's field equations. Many efforts have been done to delineate a physical theory which finds a better compliance with Mach's principle.

In the case of quasi-stationary weak fields (the so-called post-Newtonian limit), such as in the case of the Solar System, the parameterized post-Newtonian (PPN) parameters have been introduced in the expression of the metric tensor components to compare and classify alternative metric theories of gravity (see e.g. [7]). Each of these parameters has a specific physical significance and assumes a fixed value in the limit of validity of GR. Obtaining an evidence for these scalar fields' existence is an objective of paramount importance for a better understanding of the nature of space-time. The introduction of the PPN parameters paves the way to a set of fundamental physics experiments.

One of the most accurate GR tests available so far in the PPN framework is the Solar Conjunction Experiment (SCE) of NASA's Cassini mission to Saturn, carried out in 2002 [8]. This experiment aimed at the estimation of the PPN parameter $\gamma$ (the so-called Eddington parameter), which roughly controls the amount of space curvature produced by a unit mass. $\gamma$ appears in the expression of the space components of the metric tensor and a value of 1 indicates compatibility with GR. The physical effect detected by Cassini SCE was the frequency shift experienced by a radio signal exchanged between the spacecraft and an Earth station during a solar conjunction (see figure 1). This effect is due to the space-time curvature produced by the presence of a massive body as the Sun, which can be expressed as a function of $\gamma$. The Cassini SCE was able to determine that $\gamma - 1 = (2.1 \pm 2.3) \times 10^{-5}$. This measurement accuracy was obtained thanks to a multi-link configuration which provided plasma-free Doppler observables even at low Sun-Probe-Earth (SPE) angles [9], namely when the relativistic signal is stronger. The Mercury Orbiter Radioscience Experiment (MORE) on board of the BepiColombo ESA/JAXA mission to Mercury is endowed with a state-of-the-art radio tracking system with a pseudo-noise (PN) high-rate (24 Mcps) ranging code, which provides the possibility to collect accurate plasma-free range observables at low SPE, in addition to the Doppler ones. Analyzing these observables through an orbit determination (OD) process during each one of the SSCs available during the cruise phase of the mission, MORE will provide a tighter constraint on the value of $\gamma$, while searching for violations of GR. The cruise phase of an interplanetary mission is a relatively favorable situation for this kind of tests, because the non-gravitational disturbances acting on the spacecraft are limited and almost constant in time. However, during the cruise phase of the BepiColombo mission, the harsh environment of the inner Solar System will amplify non-gravitational variable effects. By comparison, the Cassini experiment was carried out when the spacecraft was at 7.2 Astronomical Units (AU), where solar radiation pressure was much smaller and the Sun aspect angle essentially constant. The BepiColombo SCE has to be carried out in a much more dynamically active environment, where the main noise source is represented by variable solar radiation pressure accelerations induced by irregular solar activity.

Due to the similar time scales of the effects, the dynamical noise due to the solar irradiance fluctuations and the relativistic time delay are significantly correlated, resulting in a difficult disentanglement of the two signals. For this reason, in this work we simulate the BepiColombo SCE under different assumptions, considering a compensation strategy for the unknown non-gravitational accelerations, to provide a more realistic estimate of the attainable accuracy in the estimation of the PPN parameter γ.

The plan of the paper is the following. In section II we introduce the physical effect behind the SCE. In section III we describe the microwave tracking system of the MORE experiment. The OD process adopted for the cruise experiment is outlined in section IV. Section V provides a comprehensive description of the simulation environment. Finally, the method of analysis and the results are reported in section VI and VII, and conclusions are given in section VIII.

## II. Relativistic effects on range and Doppler observables

Interplanetary spacecraft relativity experiments are largely based on range and Doppler observables, which, after adequate processing through OD codes, provide an estimate of dynamic parameters and physical constants. Doppler observables express the frequency shift experienced by a carrier signal generated at the ground station, received by the spacecraft, and coherently retransmitted back to Earth. Range measurements are obtained by measuring the round-trip light time $T$ between the spacecraft and a ground antenna, obtained by a suitable modulation of the carrier.

The relativistic time delay and frequency shift are a consequence of the deflection and retardation undergone by the signal during a SSC, because of the space-time curvature induced by the mass of the Sun. Thus, range and Doppler observables are consequently modified.

The PPN parameter $\gamma$ was introduced by Arthur Stanley Eddington in [10] to account for the presence of an additional scalar gravitational field. In this case, the metric tensor components of a static spherically symmetric body, can be expressed with the PPN notation as:

$$g_{00} = 1 - \frac{2GM}{c^2 R} + 2\left(\frac{GM}{c^2 R}\right)^2, \qquad (1)$$

$$g_{11} = g_{22} = g_{33} = \left(-1 + \frac{2\gamma GM}{c^2 R}\right) \qquad (2)$$

from which it is possible to retrieve the expression of the relativistic time delay (see [11], Appendix C):

$$\Delta T = \frac{(1+\gamma) G M_\odot}{c^3} \ln\left(\frac{r_1 + r_2 + r_{12}}{r_1 + r_2 - r_{12}}\right) \qquad (3)$$

where G and $M_\odot$ are the Newtonian gravitational constant and the mass of the Sun respectively, and $r_1$, $r_2$ and $r_{12}$ are shown in figure 1. When the spacecraft is close to an SSC this effect is magnified, therefore more easily detectable. In this case, it is possible to approximate the equation of the relativistic time delay and frequency shift of the signal as [7]:

$$\Delta T = \frac{(1+\gamma) G M_\odot}{c^3} \ln\left(\frac{r_1 r_2}{b^2}\right) \qquad (4)$$

$$\frac{\Delta \nu}{\nu} = \frac{d \Delta T}{dt} = -2 \frac{(1+\gamma) G M_\odot}{c^3 b} \frac{db}{dt} \qquad (5)$$

where $b$ is the impact parameter, (i.e. the minimum distance between the center of mass of the Sun and the light path of the radio wave), $\Delta T$ the relativistic time delay and $\frac{\Delta \nu}{\nu}$ the relativistic relative frequency shift in a two-way signal.

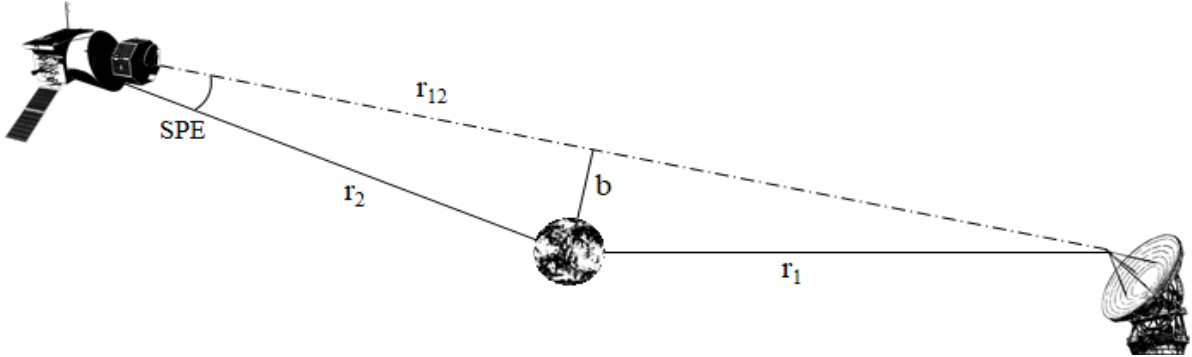

**Figure 1: Superior solar conjunction geometrical configuration. The intervening distances have to be taken at the appropriate coordinate times.**

As indicated by equations (4) and (5), the relativistic signal increases when the impact parameter is small and when its temporal rate of change is fast. The typical magnitude of the delay and the delay rate during the SSCs of BepiColombo are 66 $\mu$s and $10^{-10}$ (corresponding to a range rate of 0.03 m/s). However, in the geometrical configuration in which these two conditions are met, microwave signals travel within the solar corona environment, a region that is very harsh for their propagation, especially at lower frequencies (the refractive index of the plasma, therefore the plasma path delay, is inversely proportional to the square of the carrier frequency).

The turbulent plasma environment of the solar corona represents indeed the main disturbance for interplanetary missions' telecommunications, to the point that it is usually avoided to rely on data retrieved when the value of the SPE angle is low (usually for SPE < 35° [12]). The multi-frequency radio link of the MORE experiment, based on the simultaneous transmission and reception of X- and Ka-band signals (respectively 7.2-8.4 GHz, and 34.0-32.5 GHz) will allow a nearly complete suppression of the plasma noise, providing reliable range and Doppler data even at small SPE angles and enabling measurements of the relativistic effects near their peak.

However, when the signal passes too close to the Sun (say below 5-7 solar radii), the X-band link enters in the strong scintillation regime and the measurements become unreliable. More details are given in the next section, where the MORE radio tracking system is described.

## III. The MORE investigation on BepiColombo

The ESA/JAXA BepiColombo mission was successfully launched on 20 October 2018, from the European spaceport in Kourou, French Guiana. The Mercury Composite Spacecraft (MCS), currently in cruise to Mercury, is composed by three different modules: the Mercury Magnetospheric Orbiter (MMO), built by JAXA for the study of the magnetosphere and the exosphere of the planet, the Mercury Planetary Orbiter (MPO), developed by ESA with the aim of investigating the past surface processes and the deep interior of Mercury, and the Mercury Transfer Module (MTM), which hosts the solar electric propulsion system (SEPS) [13]. After one Earth, two Venus and six Mercury flybys, the MCS will reach Mercury on 5 December 2025, where it will execute a series of maneuvers to insert the MMO and the MPO in their respective science orbits. The MMO will separate from the MCS on 20 December 2025 to enter in a high-eccentricity orbit with a periherm altitude of 590 km and an apoherm altitude of 11639 km. On 15 March 2026, the MPO will be inserted in a low-eccentricity nearly polar orbit (480 × 1500 km altitude). The MORE experiment, one of the eleven investigations onboard the MPO, will exploit range and range-rate data to determine the gravity field, tides and rotational state of the planet Mercury and to perform a test of GR [14, 15, 16, 17]. The core of the radio science experiment is a Ka-band transponder (KaT), manufactured by Thales Alenia Space Italy, that enables a two-way coherent radio link between a ground station and the spacecraft, providing accurate Doppler and ranging measurements. Nowadays, the only deep-space stations capable of establishing a two-way link in Ka-band are ESA's DSA-3, located in Malargue (Argentina), and NASA's DSS-25, located in Goldstone (California). Both are planned to support the MORE investigation. The separation in longitude of the two stations is about 45 degrees, thus, they cannot provide two complete passes per day. The baseline tracking configuration for MORE envisages an 8-hour radio tracking pass per day. Preference to one or the other antenna will be based on the most favorable elevation profile.

The novel PN ranging system at 24 Mcps was designed to provide a single-shot precision of 20 cm after a few seconds integration time. The Doppler measurements are expected to have an Allan Deviation (ADEV, frequency stability) of $1.4 \cdot 10^{-14}$ at 1000 s of integration time, equivalent to a range rate accuracy of 0.004 mm/s [18]. Under the simplifying assumption of white noise, the accuracy on range-rate scales with the inverse square root of time and thus at 60 s it is equal to 0.017 mm/s. Recent in-flight test from ESA's DSA-3 provided significantly better results: the root mean square (rms) of range measurements is ~ 1 cm at 4 s of integration time at a distance of 0.3 AU and the range-rate measurements can be established with an accuracy in the range 0.006-0.015 mm/s at 60 s (0.015-0.030 mm/s @ 10 s), when the winds at the ground antenna do not exceed 8 m/s [19].

MORE's KaT will operate together with the Deep Space Transponder (DST) to implement a multi-frequency radio link which will guarantee this performance at almost all SPE angles. This configuration of the radio system, already adopted for range rate only by the Cassini mission, is crucial to the success of the SCE. The DST provides two downlinks at X- (8.4 GHz) and Ka-band (32.5 GHz) coherent with an uplink at X-band (7.2 GHz), while the KaT enables a Ka/Ka link (34 GHz uplink and 32.5 GHz downlink). Indeed, in the limit of geometric optics, a linear combination of the observables (range and range-rate) at the three received frequencies cancels out the plasma noise both on range and Doppler observables [8, 9]. For a coherent two-way radio link, the expression of the generic observable quantity $z$ (range or range-rate) is the sum of the plasma-free observable $z_{nd}$ and the uplink and downlink dispersive contributions:

$$z = z_{nd} + \frac{P_{up}}{f_{up}^2} + \frac{P_{down}}{f_{down}^2} \qquad (6)$$

where $P_{up}$ and $P_{down}$ are proportional to the Total Electron Content of the medium (TEC, $electrons/m^2$) for range measurements and to its time-derivative for Doppler ones, and scaled with the uplink and downlink carrier frequencies $f_{up}$ and $f_{down}$ (where $f_{down} = \alpha f_{up}$ and $\alpha$ is the transponder ratio). The purpose of the calibration scheme is to recover the non-dispersive component $z_{nd}$ from the received observable z. Thanks to the 5-link configuration, three independent observables are acquired at the ground station, and equation (6) can be written for each one, leading to a system of three equations with the three unknowns $P_{up}, P_{down}$ and $z_{nd}$. The plasma-free observable will be obtained as:

$$z_{nd} = \left(\frac{1}{\beta^2-1}\frac{\alpha_{xx}^2}{\alpha_{kk}^2}\frac{\alpha_{xk}^2-\alpha_{kk}^2}{\alpha_{xx}^2-\alpha_{xk}^2}\right)z_{xx} + \left(\frac{1}{\beta^2-1}\frac{\alpha_{xk}^2}{\alpha_{kk}^2}\frac{\alpha_{kk}^2-\alpha_{xx}^2}{\alpha_{xx}^2-\alpha_{xk}^2}\right)z_{xk} + \left(\frac{\beta^2}{\beta^2-1}\right)z_{kk} \qquad (7)$$

where $\beta = \frac{f_{upk}}{f_{upx}}$, and $\alpha_{kk} = \frac{3360}{3599}, \alpha_{xk} = \frac{3344}{749}, \alpha_{xx} = \frac{880}{749}$ are the transponder ratios.

The plasma noise cancellation scheme will fail when the impact parameter reaches a few solar radii because the effect of magnetic corrections in the refractive index and density gradients in the solar corona affects the capabilities of the scheme. Furthermore, at small impact parameters (typically at $b < 7$ solar radii) the X-band link enters in the strong scintillation regime and signal fading could occur frequently [20]. Cassini SCE verified that the cancellation scheme failed when $b < 5\ R_\odot$, and that the Allan deviation was stable at the value of $10^{-14}$ at 1000 s of integration time when $b > 7\ R_\odot$ [21]. Note that for range measurements, the absolute determination of the TEC requires a good calibration of the transponder and ground station delays, but the knowledge of its effect on the non-dispersive observable is actually enabled with very good accuracy by the onboard delay calibration system on the Ka/Ka link.

## IV. Method and orbit determination process

Precise Orbit Determination (POD) is based on the comparison of range and Doppler measurements collected by the ground stations (*observed observables*) and the ones which can be computed through an observation model (*computed observables*). Adopting a non-linear least squares (NLS) fit, the difference between computed and observed observables (*residuals*) is minimized adjusting the parameters of a selected solve-for list (e.g. initial conditions of the spacecraft orbit, forces acting in the equations of motion, physical constants). When all these parameters are accurately known or determined, the statistical properties of the residuals become compatible with those expected for the sum of the intervening noise sources. Signatures in the residuals provide evidence that the adopted dynamical or observation model is inadequate, and that refinements are needed. Note that in the case of a non-deterministic dynamical model (e.g. with unmodeled accelerations acting on the spacecraft) it may be impossible to reduce the residuals to the noise level. Such a situation needs to be carefully addressed.

Once the solve-for parameter list has been selected, the weighted NLS iterative procedure with a priori information provides an estimate of a corrected state vector which can be described by the relation [22]:

$$\delta x_{est} = (H^T W H + W_{AP})^{-1}(H^T W \delta y + W_{AP} \delta x_{AP}) \qquad (8)$$

where $x_{est}$ is the n-dimensional solve-for parameters vector, $H$ is the design matrix, containing the partial derivatives of the observables $y$ with respect to the state vector $x_{est}$ elements, $W$ is the weight matrix and $\delta x_{AP}$ and $W_{AP}$ are the a priori estimate of the state vector and its covariance matrix, respectively.

## V. Simulation setup

The numerical simulations of the SCE are based on the generation of synthetic range and Doppler data. To retrieve the simulated observed observables, we build a realistic setup (as summarized in table 1 and described in detail in sections V.1, V.2 and V.3) and then propagate a simulated trajectory (reference trajectory), starting from initial conditions of the spacecraft orbit obtained from the SPICE kernels [23] of the BepiColombo mission.

| **Observation model** | | |
|---|---|---|
| Observables | Range | Range-rate |
| Noise | 2 cm | 0.013 mm/s |
| Count-time | 300 s | 60 s |
| Earth station | DSA-3 (Malargue) | |
| Observation arc duration | 14 days across SSC | |
| **Dynamical model** | | |
| Gravity | Sun, planets, minor bodies | |
| Solar-wind | Neglected | |
| Solar radiation pressure | Modeled with SORCE/TIM solar irradiance data | |

**Table 1: Main setup features for the synthetic observables generation.**

The dynamical model used to obtain the reference trajectory is based on a number of adjustable parameters. Using an observation model, we then generate the observed observables from this trajectory. The observed observables contain also noise whose statistical properties are those expected for the MORE radiometric data [19]. The synthetic measurements are processed following a standard orbit determination procedure: a propagated trajectory is obtained from different spacecraft initial conditions and dynamical model parameters. This trajectory yields the expected value of the observables (computed observables). The estimated parameters are corrected using the iterative procedure outlined in section IV, in order to minimize a cost function, usually the weighted sum of squares of the residuals (observed observables minus computed observables). All the numerical simulations presented in this work have been carried out with the orbit determination software MONTE (Mission analysis Operations and Navigation Toolkit Environment), developed by NASA JPL [24].

**V.1 Measurements model**

The synthetic observed observables are generated considering a realistic dynamical model and the design characteristics of the tracking system: the range-rate measurements are simulated every 60 s assuming a white

Gaussian noise with rms of 0.013 mm/s; range observables are simulated every 300 s with noise of 2 cm rms. These assumptions are realistic since (as reported in section III) an analysis of first MORE's in-flight data showed that range measurements can be established with an accuracy at centimeter level after a few seconds of integration time and range-rate at 0.01 mm/s at 60 s integration time [19]. We set the spacecraft elevation threshold at the ground station to 15°. We simulate here only the data from ESA's DSA-3 tracking station, although also DSS-25 will be able to provide similar measurements. Following a conservative approach, we choose to simulate synthetic observables only when $b > 7R_\odot$, considering that if the signal path is below this threshold the plasma-noise compensation scheme cannot effectively work.

The relativistic time delay considered in MONTE is based on the formulation of Moyer [11], which is different from equation 3 because it takes into account also the second-order correction (in terms of $GM_\odot/c^2$) due to the bending of light path

$$\Delta T = \frac{(1+\gamma)GM_\odot}{c^3} ln\left(\frac{r_1+r_2+r_{12}+\frac{(1+\gamma)GM_\odot}{c^2}}{r_1+r_2-r_{12}+\frac{(1+\gamma)GM_\odot}{c^2}}\right) \qquad (9)$$

Equation 9 is an approximation of the full second-order expression of the gravitational delay, which complete formulation has been derived in [25, 26]; however, as shown in [25], equation 9 includes the second-order terms which are enhanced during SSCs, and thus, it shows to be a good approximation considering MORE sensitivity. We verified this assumption computing the numerical difference between equation 94 of [25] and equation 9 for the 6 SSCs of BepiColombo, obtaining that its maximum value is about half a centimeter (well below MORE sensitivity). Correction to post-Newtonian order 1.5, caused by the motion of the Sun, is negligible for the MORE experiment (as demonstrated in [27, 28]). A critical aspect for the predicted outcomes of the experiment is the duration of the observation arc around the SSC: in previous analyses [29, 30], a 30 days' observation arc was stated as a realistic assumption for the simulation, for each one of the eleven SSCs which happens during the cruise phase of BepiColombo.

Recent indications from the BepiColombo project limit the observation arcs to a duration of 14 days, with the possibility to increase this time span if the trajectory requirements are compliant with the flight rules. As an example, the fifth conjunction happens shortly after the third Mercury flyby and thus it would be difficult to extend the time dedicated to the SCE experiment. Furthermore, among the eleven SSCs, the SEPS will not be active only during the first 6 observation arcs (when the SEPS is active, the measurements would be useless due to the large dynamical noise). Therefore, we consider a nominal duration of 14 days for each one of the 6 observation arcs (figure 2). Although this is the baseline configuration for our analysis, we also consider how the experiment's outcome would change if:

- the threshold value for the impact parameter is smaller;

- the duration of the observation arc is extended;

- the noise on the range observables is larger.

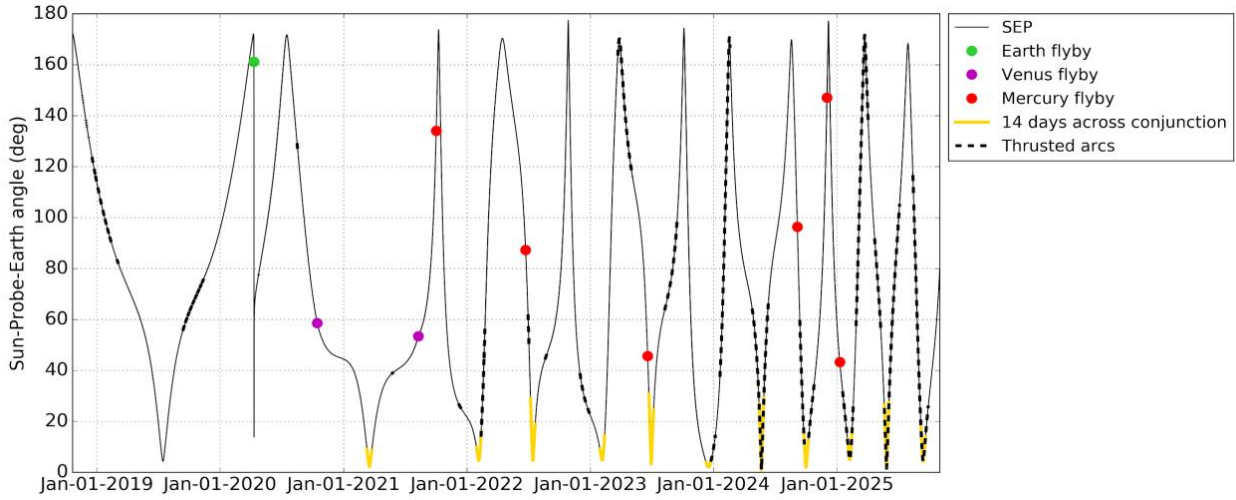

**Figure 2: The SPE angle evolution during BepiColombo interplanetary trajectory. Superior solar conjunctions occur when SPE is close to zero. The figure shows also the planetary flybys and the predicted thrusted arcs.**

## V.2 Dynamical model

The dynamical model used to generate the synthetic observables considers the gravity of the Sun, of all the planets and the main bodies of the Solar System in a relativistic formulation.

The non-gravitational acceleration produced by the solar wind can be expressed as:

$$\underline{a} = \frac{1}{2}\rho C_D \frac{A}{m} V^2 \hat{V} \qquad (10)$$

where A is the area of the exposed surfaces, m is the mass of the spacecraft, $\rho = m_p n$ where $m_p$ is the proton mass and n is the density of the charged particles ($\frac{particles}{cm^3}$), V is the velocity of the solar wind and $C_D$ is the drag coefficient. In general, low densities are associated with the fast flow originated from coronal holes, while high densities are found in the slow flow from the proximity of the streamer belt [31]. In [32] solar wind speeds were analyzed with data from OMNI and Helios for different periods between 1976 and 2006; they found out that almost all observations are in the range between 240 km/s and 750 km/s with a bimodal distribution, presenting a large slow solar wind component with a peak near 350 km/s and a smaller peak for the fast solar wind at 600 km/s. To compute the order of magnitude of the solar wind-induced accelerations, we made the standard assumption of $C_D = 2$ and, following a conservative approach, we fixed the area of the exposed surfaces to $40\ m^2$ and the mass $m = 3900\ kg$. We adopted hourly resolution data of the average velocity V and the proton density n as extracted from NASA/GSFC's OMNI data set through OMNIWeb [33]. We selected data from the period 2010-2014 to account for a realistic solar activity at the time of the experiments (considering the 11 years solar pseudo-cycle), and we found that the average value of this disturbance would be of the order of $5 \times 10^{-11}\ m/s^2$. Apart from particular events, these accelerations would lie in the range of $[1-9] \times 10^{-11}\ m/s^2$, staying close to the level of the lowest solar irradiance fluctuations (see next section). Therefore, solar wind induced accelerations would be absorbed together with the stochastic estimate of the SRP variations (see section VI). Furthermore, the space-weather-forecast-usable system anchored by numerical operations and observations (SUSANOO) [34] provides simulated solar wind parameters at the BepiColombo location. Ambiguities can be adjusted by using the in-situ measurements on BepiColombo: the sensors from the SERENA instrument package [35] and, the MIO/MPPE-MEA [36] spacecraft (in part shadowed by the sunshield) could provide valuable information on the solar wind characteristics, therefore enabling a good calibration of this effect. For these reasons, solar wind pressure was neglected in the dynamical model.

The main acceleration is due to the solar radiation pressure (SRP), the intensity of which increases as the spacecraft gets closer to the Sun. Considering a spacecraft element k, the SRP acceleration acting on its surface $A_k$ can be expressed as [37]:

$$\vec{a}_{SRP,k} = -\frac{\phi_{Sun}}{c} \frac{A_k}{m_{sc}} \left(\frac{R_{ES}}{D_s(t)}\right)^2 \left[(1 - c_{s,k})\hat{s} + 2\left(c_{s,k}(\hat{s} \cdot \hat{n}_k) + \frac{c_{d,k}}{3}\right)\hat{n}_k\right](\hat{n}_k \cdot \hat{s}) \qquad (11)$$

where $\hat{s}\ and\ \hat{n}_k$ are respectively the direction pointing from the spacecraft toward the Sun and the normal direction to the exposed surfaces, $c_{s,k}\ and\ c_{d,k}$ are the specular and diffuse reflectivity coefficients of the surfaces, $R$ is the distance from the Sun, $R_{ES}$ is equal to 1 AU, and $\Phi_{1AU}(t)$ is the solar irradiance at 1 AU from the Sun. The overall acceleration acting on the spacecraft is an integral over all the involved surface elements.

The thermo-optical coefficients' value of the surfaces together with the description of the mass' variation ([4000, 3000] kg) have been provided by the spacecraft prime contractor AIRBUS. The area exposed to the Sun depends on the attitude of the solar panels and the spacecraft bus. In our simulations, the bus is represented by a cylinder with length 6.6 m and radius of 1.8 m. We verified that the MPO solar panel is kept edge on with respect to the Sun during the interplanetary trajectory, thus it is neglected in the geometrical model of the spacecraft. The attitude of the MTM solar arrays varies during the cruise phase according to the distance from the Sun in order to avoid degradation or failure of the solar cells [38]. Therefore, in our simulations, we modified the exposed area of the panels according to the indications of [38]. With these assumptions, the intensity of the solar radiation pressure acceleration acting on the spacecraft during each observation arc will be of the order of $1 - 3 \times 10^{-7} \ m/s^2$.

**V.3 Solar irradiance**

The value of $\Phi_{1AU}$ in equation (11) shows temporal random fluctuations ranging at the level of [0.01, 0.1] % over a few hours. These oscillations of the solar irradiance are due to the simultaneous effect of faculae and sunspots on the solar surface [39]. Space-based radiometers on board of satellites orbiting in the Earth's neighborhood can measure these irradiance fluctuations with excellent accuracy; NASA's SOlar Radiation and Climate Experiment (SORCE) measured Total Solar Irradiance (TSI) fluctuations at 6-hour rate with an accuracy of 350 ppm [40]. To generate the observed observables, we included in the dynamical model the irradiance fluctuations measurements of the SORCE mission (figure 3); these data are taken every 6 hours [41], and represent the value of TSI at 1 AU from the Sun. We verified that even representing irradiance fluctuations with the higher cadence data of SOHO [42] (SOlar Heliospheric Orbiter) mission dataset (one TSI measurement per minute, [43]) the outcomes of the simulation remain unchanged.

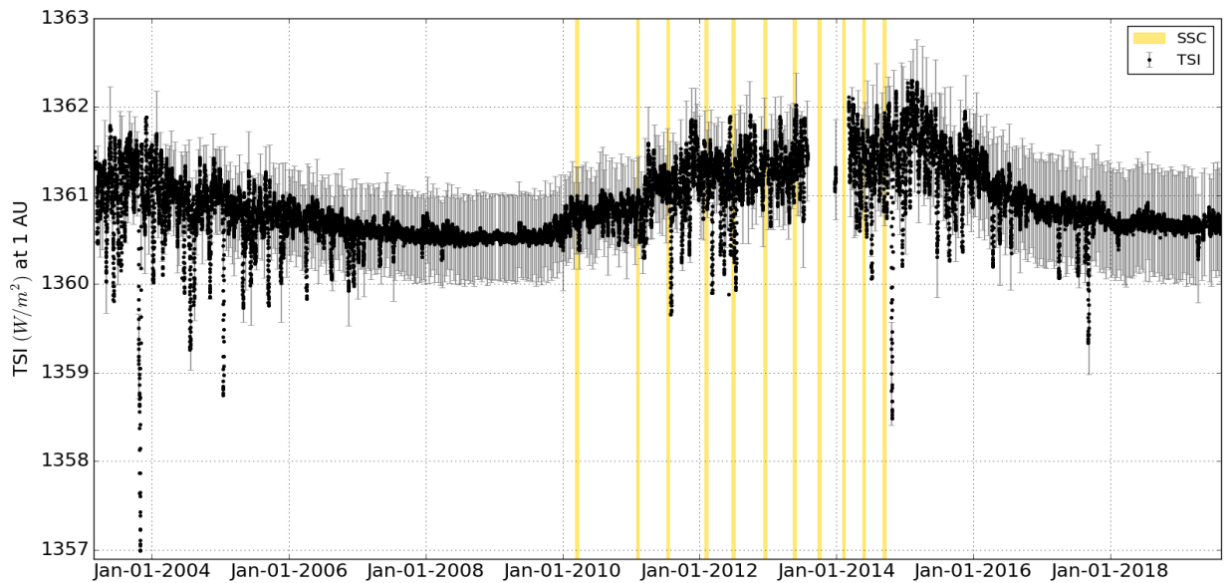

**Figure 3: Total solar irradiance measured by SORCE in the last 11 years. The yellow shaded area represents the BepiColombo superior solar conjunctions shifted by 11 years.**

Anyway, the geometry inherent in the SSC will hinder the possibility to use these measurements to calibrate the adverse effect of the irradiance fluctuations. Indeed, the faculae and sunspots responsible for the fluctuations affecting the spacecraft are located on the opposite face of the Sun from the one measured by SORCE. Therefore, the solar pressure acceleration will be predictable only to 0.1-0.01% depending upon the solar activity at the time of the experiment. The MPO hosts a three-axis high-sensitivity accelerometer (the Italian Spring Accelerometer, ISA) [42] with the aim of measuring non-gravitational accelerations acting on the spacecraft during the orbital phase [44]. The largest SRP-induced accelerations have characteristic frequencies in the range $[10^{-6}, 10^{-5}]\ Hz$, outside the operational range of ISA, $[3 \times 10^{-5}, 10^{-1}]\ Hz$. Furthermore, the faint signal induced by SRP fluctuations is at the level of $10^{-10}\ m/s^2$, well below the instrument sensitivity (about $10^{-8}\ m/s^2$) [45] (a detailed discussion of this issue can be found in [46]).

The dynamical noise induced by the solar irradiance fluctuations is a major concern for the SCE because the signal induced by SRP fluctuations, although a small fraction of the relativistic signal (about 0.01% in delay and 0.07% in delay rate over 24 h), may reduce the accuracy in the determination of the PPN parameter $\gamma$. Neglecting this effect would lead not only to an illusorily tight constraint on $\gamma$, but also to a dangerous bias in the estimation. We therefore modify the dynamical model assuming a constant solar irradiance (varying as $1/R^2$), and compensate with stochastic accelerations the dynamical noise induced by the irradiance fluctuations.

## VI. Stochastic estimation of solar irradiance

### VI.1 Method

The computed observables are generated according to a trajectory based on the dynamical model described in section V, but with a constant solar irradiance equal to 1360.5 $W/m^2$. In this way, we represent the lack of knowledge of the SRP fluctuations and the dynamical environment during the experiment. We verified that uncalibrated fluctuations of the solar irradiance (figure 4, top) generate noticeable signatures on the range residuals (figure 4, bottom), proving the inadequacy of a deterministic dynamical model. In this case the estimated parameters are:

- The spacecraft state vector
- The Eddington parameter $\gamma$ (with an a priori uncertainty equal to the outcome of Cassini SCE)
- A constant scale factor for the average solar irradiance value
- One range bias in each observation arc (with an a priori uncertainty equal to 6 m).

We verified that the unmodeled spacecraft dynamics introduces a bias in all these parameters that is three times larger than the formal uncertainty.

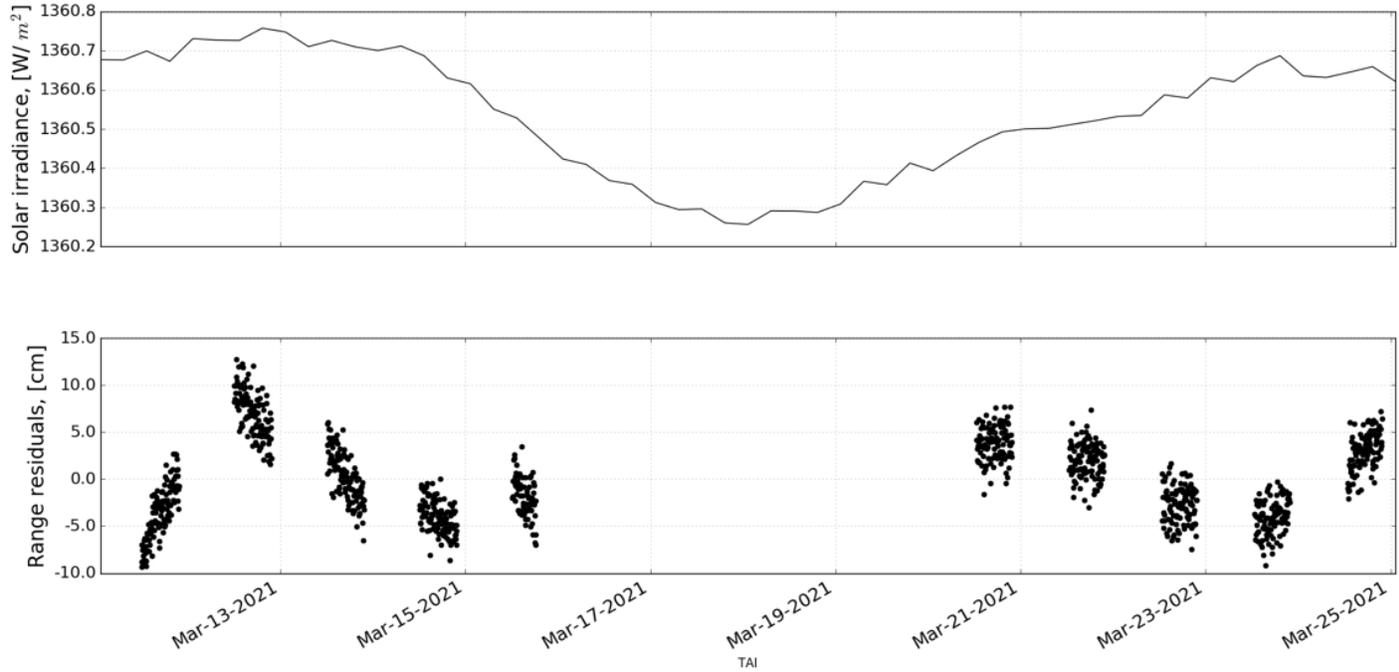

**Figure 4: Fluctuating solar irradiance pattern adopted in the simulation of the radiometric measurements (top) and the resulting range residuals after 3 iterations (bottom), when no compensation strategy is adopted to absorb the effect of the unknown SRP acceleration fluctuations.**

To compensate for this partially unknown acceleration we extended the solve-for parameters list by adding a variable irradiance scale factor which can vary in different time batches. We selected an exponentially correlated random variable (ECRV) model to take into account the time correlation of the solar fluctuations [16].

The time evolution of the ECRV from one batch to another is ruled by the *Langevin* differential equation:

$$\dot{\eta}(t) = -\frac{1}{\tau}\eta(t) + u(t) \tag{12}$$

The right-hand side of equation (12) underlines that the scale factor time variation depends on a white noise contribution $u(t)$, but also on a deterministic term controlled by a time constant $\tau$.

The ECRV stochastic model of the scale factor is described by the following parameters, to be specified as inputs of the estimation process:

- The uncertainty $\sigma_{st}$, which describes the stochastic parameter variance.
- The batch time duration $\Delta t$, which depends on the predicted timescale of the stochastic phenomenon.
- The time constant $\tau$, correlation time between parameters in different batches.

The correlation time $\tau$ is set to 27.2753 days, that is the solar rotation period [47]. This indication comes from the analysis of 11 years of SORCE TSI measurements, which shows an autocorrelation peak at the Sun rotational period.

The choice of $\sigma_{st}$ and of the batch time $\Delta t$ depends upon the solar activity during the conjunction. Analyzing the measured TSI for the 11-year pseudo-cycle, it appears that during the period of solar maxima the activity is stronger and irregular (reaching fluctuations of the 0.1% of the central value), while during the minima the irradiance fluctuations standard deviation is lower (beneath the 0.01%). Shifting the solar activity by eleven years, it is expected that most SSCs will happen during strong solar activity periods. Anyway, during these periods the irradiance fluctuations behavior varies between intervals of strong and weak fluctuations, and thus the possibility that these oscillations are in the lower end of the range is admitted.

In this paper we assume that there is no possibility to predict precisely the specific frequency and standard deviation of the SRP variations. We therefore search empirically for the optimal combination of $\sigma_{st}$ and $\Delta t$. The technique that we adopted for the simulations is based on the analysis of the range residuals' distribution obtained by varying the combination of $\sigma_{st}$ and $\Delta t$. If this choice is inappropriate with respect to the characteristics of the simulated solar activity, the range residuals' distribution would present strong signatures. This allows us to recognize an unsuitable combination of noise model parameters and try other values. Among the admissible combinations (i.e. those which result in a white and Gaussian distribution of the range residuals), we select the $\sigma_{st}$ and $\Delta t$ which provide the best accuracy on $\gamma$.

This technique will lead to a realistic indication on the SCE outcome. However, it also implies a large number of simulations varying the input $\sigma_{st}$ among all the possible values compliant with all kinds of fluctuations' intensity. In order to reduce the computational burden, it is useful to start from a realistic magnitude of the irradiance fluctuations. Performing a first step analysis with a large value $\sigma_{st}$ and a tight $\Delta t$ it is possible to retrieve information on the standard deviation of the SRP oscillations. In this case, we are not aiming at the estimate of $\gamma$, but only at a posteriori reconstruction of solar irradiance. Figure 5 illustrates an example of the estimation of a TSI pattern. This information allows a tightening of the range of $\sigma_{st}$ to be considered for the selection procedure.

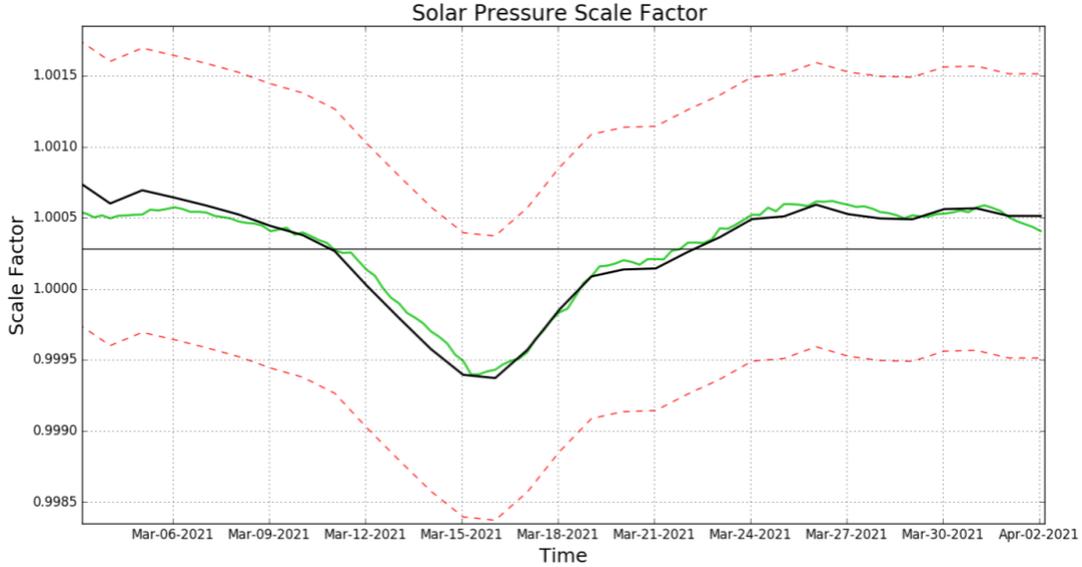

**Figure 5: Estimation of solar irradiance fluctuations: the green line represents the simulated SRP model while the black line is the one reconstructed with a batch time of 24 h and a conservatively large a priori uncertainty of $1 \times 10^{-3}$. The red dashed lines represent the three-sigma interval.**

**VI.2 Discussion**

The value of $\sigma_{st}$ gives an indication of how much the estimated scale factor can deviate in each time interval from the constant a priori value: this means that a tight $\sigma_{st}$ prevents the possibility to estimate an irradiance value farther than $\sigma_{st}$ from the a priori. Consequently, the residual effect of the SRP acceleration causes a bias on the estimated value of $\gamma$ and on the other solve-for list parameters, with signatures in the residuals' distribution. On the other hand, selecting an unnecessary large value of $\sigma_{st}$ would unrealistically increase the uncertainty of the estimation of $\gamma$: the two signals have a similar effect on the observables, therefore a larger uncertainty on the estimated irradiance scale factor reflects as a larger uncertainty on $\gamma$. The same principles describe the selection of the batch time duration $\Delta t$: when the batch time is too large with respect to the actual rate of change of the TSI, the effect of the variation experienced at higher frequency would be neglected, amplifying the bias on the estimated parameters. At the same time, a shorter batch time adds a large number of scale factors to the solve-for list of the POD process, increasing the uncertainty on $\gamma$ and the other estimated parameters.

## VII. Results

### VII.1 Covariance analysis

If precise measurements of the irradiance fluctuations were available to model the dynamic environment during the experiment, the accuracy attainable on the estimation of γ would be dictated only by the limitations of the tracking system. It is worth to point out which would be the constraint achievable on γ if the dynamical model of the spacecraft was perfectly known, which acts as a best case for the experiment results. In this case, we include in the solve-for list the state vector of the spacecraft, the PPN parameter γ (with an a priori uncertainty fixed by the Cassini SCE), a constant scale factor for the average value of the solar irradiance, and then we consider two cases for the range biases. In table 2 the uncertainties related to the determination of γ are indicated under four different assumptions:

1. setting the noise of the range observables at 2 cm and estimating one single range bias per arc
2. keeping the noise of the range observables at 2 cm and estimating one range bias per pass
3. increasing the noise of the range data at 20 cm and estimating one single range bias per arc
4. same assumptions of case 1 but extending the duration of the observation arc to 30 days.

|        | SSC 1 | SSC 2 | SSC 3 | SSC 4 | SSC 5 | SSC 6 |
|--------|-------|-------|-------|-------|-------|-------|
| Case 1 | 0.5   | 0.3   | 0.16  | 0.44  | 0.18  | 1.8   |
| Case 2 | 1.6   | 1.3   | 1.0   | 1.6   | 0.99  | 2.1   |
| Case 3 | 1.9   | 1.6   | 1.1   | 1.9   | 1.1   | 2.2   |
| Case 4 | 0.078 | 0.070 | 0.073 | 0.086 | 0.064 | 0.34  |

**Table 2: Covariance analysis uncertainty on the estimation of $\gamma$ ($\times 10^{-5}$): 1) with 2 cm noise on range data, estimating one range bias during the whole observation arc; 2) with 2 cm noise on range data, estimating one range bias each pass; 3) increasing the noise on range data to 20 cm, estimating a single range bias per arc; 4) as in case 1 but with an extension of the observation arc up to 30 days.**

Our baseline assumptions are represented by case 1: as table 2 points out, the performance of the range measurements is a key factor for the success of the experiment (i.e. a precise determination of the Post-Newtonian parameter γ); in case 3, the increase of the estimation uncertainty on γ with respect to case 1, ranges from a minimum of 22% in SSC #6 to a maximum of 85% in SSC #3. Retrieving range and Doppler observables for 30 days would give the possibility to detect the relativistic signal in a region where the space curvature has decreased significantly. Although Doppler data alone would not provide a noticeable gain, the range observables proved to be particularly sensitive to this additional information, reaching the limit of $\sigma_\gamma = 6.4 \times 10^{-7}$ with SSC #5 alone. The increase of estimation uncertainty experienced in case 2 is due to the addition of a large number of parameters to the solve-for list (one range bias per pass): estimating more parameters with the same amount of information leads to an unavoidable increase of the uncertainties on the solve-for list elements and interrupts the coherence of the measurement over a multi-pass timespan. The possibility to estimate only one range bias per arc was suggested by the analysis of the first MORE in-flight data of the novel PN range at 24 Mcps [19], but further tests will be performed in September-November 2020 to verify this result in a cleaner dynamical environment.

### VII.2 Nominal assumptions results

The possibility to get a satisfactory constraint on the value of the Eddington parameter is closely related to the level of the solar activity during the experiment. For this reason, we repeated for each conjunction the simulation of the SCE varying the SRP pattern adopted to generate the synthetic observables. Considering the 11 years solar pseudo-cycle, we shift the epochs of each SSC at the time of the actually available SORCE measurements, and then we simulate 50 irradiance fluctuations models through data collected in a two-year period of time centered in the SSC (as shown in figure 6 for SSC #1).

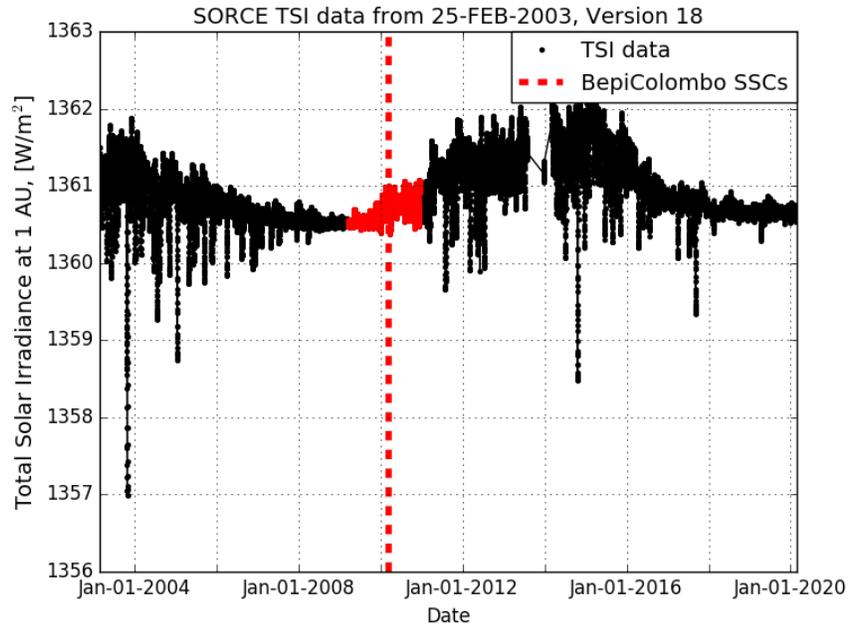

**Figure 6:** Total solar irradiance measured by SORCE in the last 11 years. The red data have been exploited to simulate the SRP models used to generate synthetic observables for SSC #1.

This choice is driven by the need to avoid simulating unlikely levels of solar activities with respect to those corresponding to the period of the experiment. By varying the simulated SRP pattern, the uncertainty with which γ can be estimated will change. Therefore, we report for each conjunction the average of the results obtained from 50 simulations, and the tightest value corresponding to the possible best case. In the worst case scenario (corresponding to the harshest solar activity), no improvement would be obtained over the Cassini result [8].

With our baseline assumptions about the data collection (14 days arcs' duration, 2 cm of range measurements' noise, and excluding data retrieved when the impact parameter is smaller than 7 solar radii), only SSCs #1, #3 and #5 are able to exceed a limit of $2 \times 10^{-5}$ in the formal uncertainty on γ. In table 3 we indicate the results attainable by considering a single SSC. Performing a combined analysis which exploits a joint dataset of the three conjunctions, the PPN parameter γ can be determined to an accuracy of $1.3 \times 10^{-5}$. With this same approach but considering the best cases indicated in table 3, the attainable uncertainty would be reduced to $6 \times 10^{-6}$.

|  | Average result ($\times 10^{-5}$) | Best case ($\times 10^{-5}$) |
|---|---|---|
| SSC #1 | 1.9 | 0.73 |
| SSC #3 | 1.9 | 0.71 |
| SSC #5 | 1.8 | 1.14 |

Table 3: Uncertainty on the estimation of γ considering baseline assumptions. The first column describes the average of the results obtained with the 50 simulations, while in the second column the best cases are reported.

### VII.3 Sensitivity analysis

As stated in section III, the applicability of the plasma noise cancellation procedure depends on the value of the impact parameter. The quality of the data deteriorates gradually when the signal path digs deep into the solar corona, therefore we choose to consider a perfectly working plasma calibration until $b_{CUT} = 7 R_{SUN}$ and to ignore data collected below this threshold. At the same time, the outcome of the test can be improved if the cut-off value of the impact parameter would be lower. As a precise prediction to understand which is the limit of operation of the plasma compensation scheme is not available, it is worthwhile to explore the outcome of the experiment when the multifrequency plasma compensation scheme is effective at smaller impact parameter thresholds. Figure 7 represents the formal uncertainty in the estimate of γ attainable in each SSC as a function of the threshold impact parameter. SSCs #2 and #4 have a minimum impact parameter that is larger than 7 solar radii, thus a reduction of the cut-off value has no effect on their SCE outcome. The best result attainable with a single conjunction is in the case of a weak solar activity with $b_{CUT} = 5 R_{SUN}$ for the first SSC (March 2021), reaching an accuracy of $6 \times 10^{-6}$. The same SSC is able to obtain a similar constraint on γ even if $b_{CUT} = 6 R_{SUN}$.

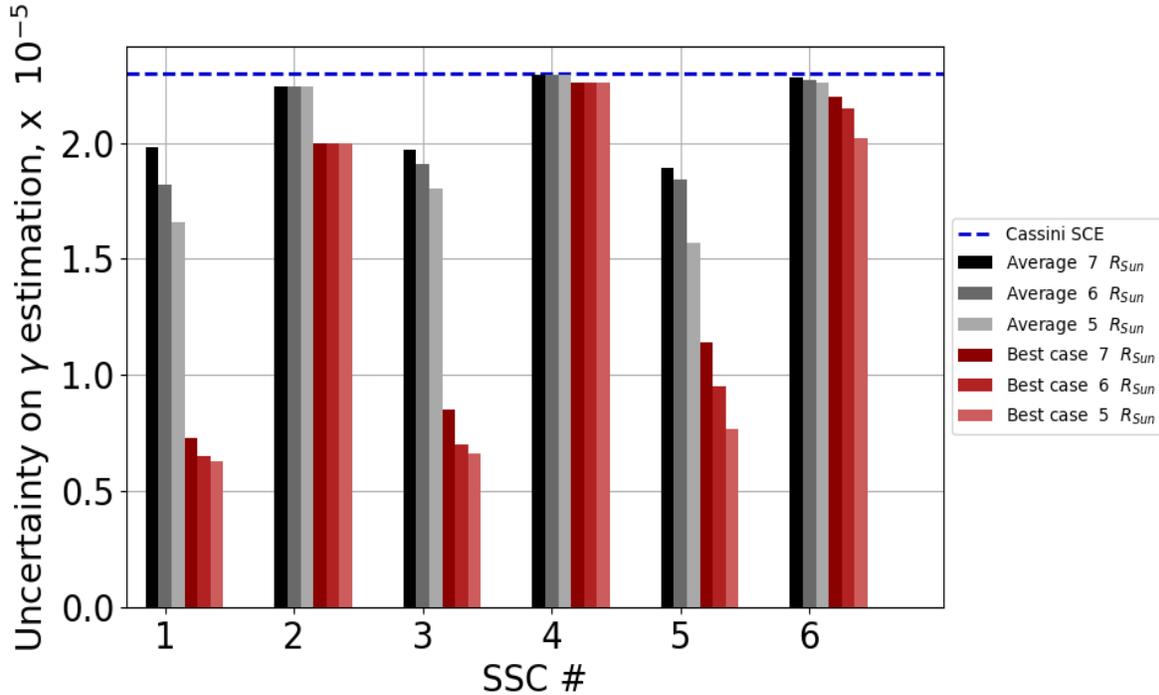

Figure 7: Uncertainty on $\gamma$ for different threshold values of the plasma noise calibrations.

An extension of the observation arc duration allows, in general, to better constrain the trajectory across the solar conjunction and to reduce the uncertainties in all estimated parameters (see table 2). However, when taking into account the random unmodeled non-gravitational perturbation caused by the variable TSI, an improvement of the SCE result as a consequence of a wider observation window is not obvious: in case of medium to large SRP fluctuations, even extending the observation arc duration to 30 days [29, 30], the outcome of the experiment would not be considerably improved. On one hand, a larger observation window would provide for additional data to detect the relativistic signal, on the other hand, the unknown SRP fluctuations are acting for a longer time, accumulating the effect on the observables. In the case of a favorable SRP model, a variable observation window duration would lead to different results (figure 8). It is worth noticing that some SSCs show an increase in the formal uncertainty in $\gamma$ rather than the expected improvement shown in table 2. This behavior is due to the fact that a SRP oscillation model free from intense fluctuations for a longer time is uncommon.

However, if the observation window is entirely free from any severe TSI oscillation, the accuracy attainable on $\gamma$ could be considerably better with a longer dataset. With an extension to 20 days, SSCs #2 and #4 become more valuable opportunities. Therefore, an extension of the observation window, when possible, would be an important factor to augment the probability of a tighter constraint on the Eddington parameter.

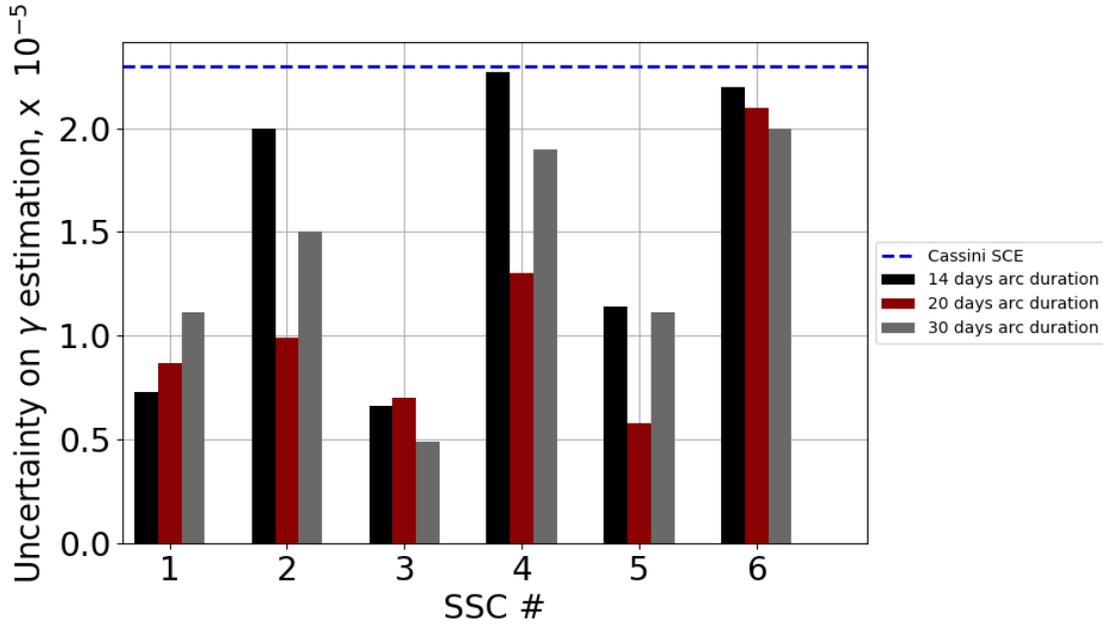

**Figure 8: Variation of the formal uncertainty on γ with an extension of the observation window.**

We also verified that, because of the presence of irradiance fluctuations, having a larger noise on range measurements (20 cm with 300 s of integration time) would not lead to a significant deterioration of the solution in the case of a high level of solar activity. Considering the best cases, the effect of the degraded accuracy in range is more noticeable: for SSCs #1, #3 and #5 the formal uncertainty on $\gamma$ increases respectively by 168%, 64%, and 3.5 %.

## VIII. Conclusions

In this work a full set of numerical simulations of the BepiColombo SCE was performed. We considered several assumptions on the duration of the observation arc, the ranging system performances and the level of solar activity affecting the irradiance fluctuations. We verified that according to the latest mission scenario, with a guaranteed observation time of 14 days across each superior solar conjunction and an average solar activity, the best constraint attainable regarding the determination of $\gamma$ would be $1.3 \times 10^{-5}$, exploiting data from SSC #1, #3 and #5. If the solar conditions are favorable, the uncertainty could be lowered to $6 \times 10^{-6}$. With these assumptions, the other three exploitable conjunctions (SSC #2, #4, #6) would not contribute significantly to the measurement of $\gamma$. However, we have found that an extension of the observation arc to 20 days would bring significant gains to SSC #2 and SSC #4 (figure 8). We also assessed the effect of the availability of good plasma calibrations at smaller impact parameters for each one of the three best conjunctions. It should be remarked that SSC #1 happens during a solar minimum, therefore it is plausible to consider a weak solar activity, and a full operation of the plasma noise cancellation procedure until 6 solar radii: in this case, thanks to the conjunction of March 2021 alone, $\gamma$ would be determined with an accuracy of $6 \times 10^{-6}$. However, if suitable models to predict solar irradiance fluctuations on the hidden face of the Sun will be available (e.g. via correlations with measured acoustic modes on the visible side), the outcome of the experiment can be substantially improved: the more accurately these predictive models can describe the irradiance fluctuations, the closer the results will get to the ones reported in table 2.

## IX. Acknowledgments

The research presented in this work has been carried out at Sapienza University of Rome and has been funded by the Italian Space Agency within the scope of the contract n. 2017-40-H.1-2020.